\documentclass[showpacs,preprintnumbers,english,twocolumn,amsmath,amssymb]{revtex4}
\usepackage[T1]{fontenc}
\usepackage[latin1]{inputenc}
\usepackage{graphicx}
\usepackage{amssymb}
\usepackage{verbatim}
\usepackage{epic,eepic}
\usepackage{babel}
\usepackage{color}
\begin{document}

\def\Journal#1#2#3#4{{#1} {\bf{#2}}, {#3} (#4).}
\def\ANP{{\em Adv. Nucl. Phys.}}
\def\ARNPS{{\em Ann. Rev. Nucl. Part. Sci.}}
\def\CTP{{\em Commun. Theor. Phys.}}
\def\EPJA{{\em Eur. Phys. J. A}}
\def\IJMPE{{\em International Journal of Modern Physics E}}
\def\JCHP{{\em J. Chem. Phys.}}
\def\JCP{{\em Journal of Computational Physics}}
\def\JPCS{{\em Journal of Physics: Conference Series}}
\def\JPG{{\em J. Phys. G: Nucl. Part. Phys.}}
\def\NATURE{{\em Nature}}
\def\NC{{\em La Rivista del Nuovo Cimento}}
\def\NCA{\em IL Nuovo Cimento A}
\def\NPA{{\em Nucl. Phys.} A}
\def\NST{{\em Nuclear Science and Techniques}}
\def\PAN{\em Physics of Atomic Nuclei}
\def\PHY{{\em Physics}}
\def\PRA{{\em Phys. Rev.} A}
\def\PRC{{\em Phys. Rev.} C}
\def\PRD{{\em Phys. Rev.} D}
\def\PLA{{\em Phys. Lett.} A}
\def\PLB{{\em Phys. Lett.} B}
\def\PLD{{\em Phys. Lett.} D}
\def\PRL{\em Phys. Rev. Lett.}
\def\PL{{\em Phys. Lett.}}
\def\PREV{\em Phys. Rev.}
\def\PREP{\em Physics Reports}
\def\PROG{{\em Progress in Particle and Nuclear Physics}}
\def\RPP{{\em Rep. Prog. Phys.}}
\def\RDNC{{\em Rivista del Nuovo Cimento}}
\def\RMP{{\em Rev. Mod. Phys}}
\def\SCIENCE{{\em Science}}
\def\ZPA{{\em Z. Phys. A.}}

\def\ANN{{\em Ann. Rev. Nucl. Part. Sci.}}
\def\ANNAST{{\em Ann. Rev. Astron. Astrophys.}}
\def\AP{{\em Ann. Phys}}
\def\APJ{{\em Astrophysical Journal}}
\def\APJS{{\em Astrophys. J. Suppl. Ser.}}
\def\EJP{{\em Eur. J. Phys.}}
\def\LANC{{\em Lettere Al Nuovo Cimento}}
\def\NCA{{\em Nuovo Cimento} A}
\def\PHYS{{\em Physica}}
\def\NP{{\em Nucl. Phys}}
\def\MATH{{\em J. Math. Phys.}}
\def\JPAM{{\em J. Phys. A: Math. Gen.}}
\def\PRO{{\em Prog. Theor. Phys.}}
\def\NPB{{\em Nucl. Phys.} B}

\title{Density and Temperature in Heavy Ion Collisions: A Test of Classical and Quantum Approaches}
%\footnote{Funded in part by DOE and NSF-REU Program}}
\author{ H. Zheng$^{a,b)}$, G. Bonasera$^{a,b)}$, J. Mabiala$^{a)}$, P. Marini$^{c)}$ and A. Bonasera$^{a,d)}$}
%\email{ruslan.magana@nucleares.unam.mx}
\affiliation{
a)Cyclotron Institute, Texas A\&M University, College Station, TX 77843, USA;\\
b)Physics Department, Texas A\&M University, College Station, TX 77843, USA;\\
c)  Centre d'Etudes Nucl\'eaires de Bordeaux Gradignan, Chemin du Solarium, Le Haut Vigneau, BP 120 F-33175 GRADIGNAN, France;\\
d)Laboratori Nazionali del Sud, INFN, via Santa Sofia, 62, 95123 Catania, Italy.}
%\author{Hua Zheng$^a)$}
%\email{hzheng@comp.tamu.edu}
%\affiliation{Cyclotron Institute, Texas A\&M University, College Station, TX 77843, USA}
%\author{Aldo Bonasera$^{a,c)}$}
%\email{abonasera@comp.tamu.edu}
%\affiliation{Cyclotron Institute, Texas A\&M University, College Station, TX 77843, USA}
%\affiliation{Laboratori Nazionali del Sud, INFN, via Santa Sofia, 62, 95123 Catania, Italy.
%\date{September 8, 2010}
\begin{abstract}
Different methods to extract the temperature and density in heavy ion collisions are compared using a statistical model tailored to reproduce many experimental features at low excitation energy.
The model assumes a sequential decay of an excited nucleus and a Fermi gas entropy. We first generate statistical events as function of excitation energy but stopping the decay chain at the first step.
In such a condition the `exact' 
model temperature is determined from the Fermi gas relation to the excitation energy. 
From these events, using quantum and classical fluctuation methods for protons and neutrons, we derive temperature and density (quantum case only) of the system under consideration.
  Additionally, the same quantities are also extracted using the double ratio method for different particle combinations. 
A very good agreement between the ``exact'' temperatures and quantum fluctuation  temperatures is obtained, the role of the density is discussed. 
Classical methods give a reasonable estimate of
the temperature when the density is very low as expected. The effects of secondary decays of the excited fragments  are discussed as well.
\end{abstract}

\pacs{ 25.70.Pq, 24.60.Ky, 64.70.Tg, 05.30.Jp}

\maketitle
%\section{Introduction}

The nuclear equation of state (NEOS) is one of the most challenging open problems today in particular the access to
 the symmetry energy part which carries relevant information especially for the nuclear (astro)physics domain\cite{baoanrep08, huappnp, baranrep05, skyrme3, nsob12}. 
A feasible way to experimentally constrain the NEOS is through the use of heavy ion collisions (HIC) at intermediate energies involving nuclei with a large range of N/A concentrations. 
The systems created in such conditions
are dynamical and strongly influenced by Coulomb, angular momentum 
and other effects, thus the determination of `quasi-equilibrium' conditions is rather challenging. 
The NEOS can be determined if we are able to extract the temperature, density and 
pressure, or free energy from the HIC data. Several methods can be found in the literature to determine such quantities from available experimental data. 
Classical approaches include slope temperature from the kinetic energy distribution
of emitted ions \cite{aldonc00, joe3, joeepja}, excited level energy distributions \cite{msuld} and double isotopic ratios \cite{albergot, aldonc00, joe3, joeepja,  msuld, msu1, msu2, qinprl12, hagelprl12, wadaprc12}. 
In particular the last method provides information on the density $\rho$ of the system at the time when fragments are emitted
from a source at a given excitation energy per particle $\frac{E^*}{A}$. A coalescence approach \cite{ qinprl12, hagelprl12, wadaprc12, mekjian1, mekjian2, awes, csernai}
can also be used to estimate the density of the system. Even though such a method is 
derived from classical physics, the obtained densities are higher than that of the double ratio method \cite{qinprl12, hagelprl12, wadaprc12, ropke}. This might be due to the introduction of a new parameter
(the coalescence radius $P_0$) which is determined from experimental data. Such a parameter might in fact mimic  important quantum effects \cite{ropke} which result in a generally higher density, for a given temperature $T$, as compared to the one obtained from the double ratio method. More recently, a different method to extract $T$ and $\rho$ from the data has been proposed in \cite{wuenschel1, hua3, hua4, hua5, hua6, hua7} based on fluctuations. The method was first used for quadrupole fluctuations
within a classical approach, in order to obtain the temperature of the system. Later on this method was applied to quantum systems for which the temperature
 is naturally connected to the density. In such a scenario, particle multiplicity fluctuations
are used in order to pin down $T$ and $\rho$ from experimental data and modeling \cite{ hua3, hua4, hua5, hua6, hua7, comd1, comd2, justinT, brianT}. 
An important first distinction between Fermions and Bosons is necessary in order to evidence important quantum effects such as Fermion quenching (FQ)
and Bose-Einstein Condensation (BEC) \cite{brianT, esteve1, muller1, sanner1, westbrook1, palao}. A correction due to Coulomb effects for Bosons and Fermions was also introduced in refs. \cite{huappnp, hua6, hua7}.
\begin{figure}
\centering
\includegraphics[width=1.0\columnwidth]{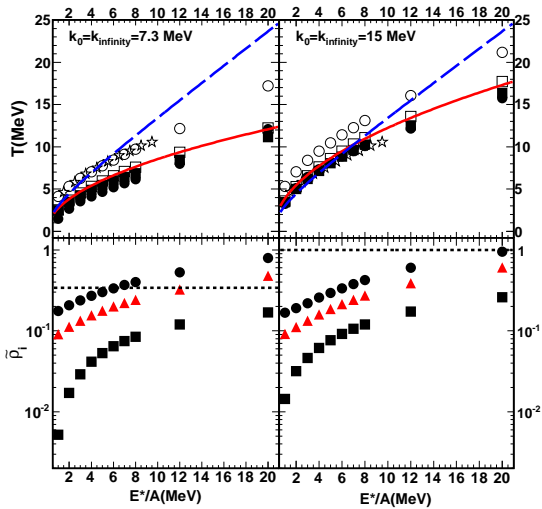}
\caption[]{(Color online) Temperature (top panels) and  density (bottom panels) as function of the excitation energy per particle.
 Different values of  $k_0$ are used in the left and right hand side panels. The exact value of $T$
obtained from the Fermi gas relation, eq. (\ref{energy}), is given by the full (red) lines. 
Available experimental data from the current literature are given by the (blue) dashed line \cite{elliott1, elliott2, elliott3} and
open stars \cite{justinT}. Open (full) circles and open (full) squares refer to the classical (quantum) fluctuation method for protons and neutrons respectively. 
The full upward triangles refer to the total density. The dotted lines refer to the total densities estimated from level density for different $k_0$, eq. (\ref{leveldens}).}\label{Fig1}
\end{figure}

In order to distinguish among different approaches and test their region of validity, we have applied the double ratio method, the classical and quantum fluctuation methods to analyze `events' obtained
 from a commonly used statistical model dubbed as GEMINI \cite{geminicode, gemini1, gemini2, gemini3, gemini4, gemini5, gemini6, gemini7}.
Similar studies using the slope temperature have been reported in refs. \cite{magemini1, magemini2}.
 The model assumes a sequential statistical decay of a hot source of mass ($A$) and charge ($Z$), with $\frac{E^*}{A}$ and a given total angular momentum $J$, which we  assume equal
 to zero for simplicity in this work. We   fix $A=80$ and $Z=40$ also in order to compare 
to many calculations based on the Constrained Molecular Dynamical model (CoMD) 
which we have performed before \cite{hua3, hua4, hua5, hua6, hua7, comd1, comd2}. The statistical model assumes that
a hot source decays into a small fragment $(A', Z')$ and a daughter  nucleus $(A-A',Z-Z')$. In general both fragments can be still excited and decay again into other fragments and so on until all excitation energy is 
transformed into kinetic energy of fragments and the Q-value determined from the initial source and the final fragments.
 At each decay step, the probability of the process is determined from the entropy which is assumed to be that
of a Fermi gas \cite{landau, huang}:
\begin{equation}
S=2aT, \label{entropy}
\end{equation}
corresponding to an excitation energy
\begin{equation}
E^*=aT^2. \label{energy}
\end{equation}
Both equations can be derived from a simple low temperature approximation of a Fermi gas. The level density parameter $a$ in such  approximation is given by:
\begin{equation}
a=\frac{A}{k_0},\label{level}
\end{equation}
for ground state density $k_0=15$ MeV.  
In order to take into account experimental observations, the parameter $k_0$ in the model is adjusted to a smaller value which could depend on excitation energy as well \cite{geminicode, gemini1, gemini2, gemini3, gemini4, gemini5, gemini6, gemini7, elliott1, elliott2, elliott3}. For our purposes
we will use two fixed $k_0$ values of 7.3 MeV and 15 MeV since our goal is to test different methods to determine $T$ and $\rho$ from the model data. 
In fact, in the model, the temperature can be derived from eq. (\ref{energy})  if we stop the  simulation after
the first decay step. The following steps take into account a decreasing temperature due to particle emissions in previous decays, thus the $T$ determination becomes  `tricky' and we will discuss it later on in the paper.
The model does not implicitly assume any density, even though one might naively think that the density of the system is that of the excited nucleus in its ground state density. 
This is not however correct since the Fermi gas relation is assumed.
In particular the level density is connected to density by the relation:
\begin{equation}
a=\frac{A}{15}(\frac{\rho}{\rho_0})^{-2/3}. \label{leveldens}
\end{equation}
\begin{figure}
\centering
\includegraphics[width=1.0\columnwidth]{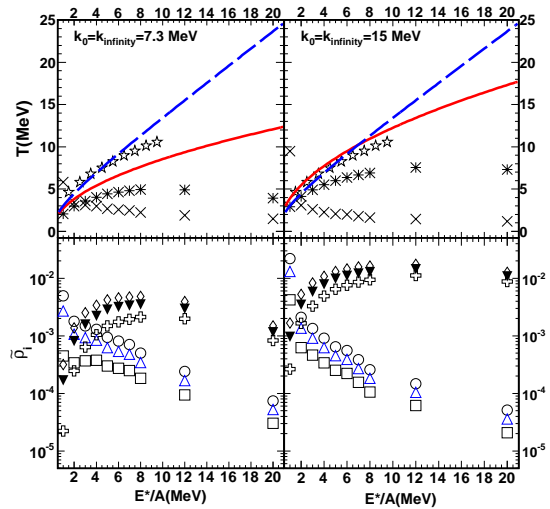}
\caption[]{(Color online) Same as fig.\ref{Fig1}. (Top panels) The asterisks and crosses refer to the $T$ obtained from DR using dth$\alpha$
and pnth combinations respectively. (Bottom panels) Open diamonds, open crosses and full downward triangles refer to p, n and total densities obtained from the dth$\alpha$ DR. Open circles, squares and upward triangles refer to the same quantities from the pnth DR.}\label{Fig2}
\end{figure}
Thus $k_0=15$ MeV implies a nuclear ground state density, while 
$k_0=7.3$ MeV results in  $\frac{\rho}{\rho_0}=0.34$. We stress here that other effects can be invoked to justify the smaller value of $k_0$, such as an effective mass or momentum 
dependence of the NEOS \cite{geminicode, gemini1, gemini2, gemini3, gemini4, gemini5, gemini6, gemini7}; however in the simple Fermi gas used in the model, the eq. (\ref{leveldens}) is the natural explanation. This has an important consequence and in fact, we would naively expect that the different approaches to
determine the density will give values compatible to the estimate above, at least asymptotically i.e. at high $T$ where different effects, such as Coulomb barriers and Q-values become less important. At low $T$, these effects are important
and will determine  the probability of decay into a given channel rather than another. We anticipate that the decay probability at low $T$ will effectively decrease the  values of the densities obtained from all methods,
which might correspond to the fact discussed in CoMD simulation, that the emitted particles at low $T$ are located in a low density region of the nuclear surface: a low $T$ will correspond to a lower $\rho$. 
\begin{figure}
\centering
\includegraphics[width=1.0\columnwidth]{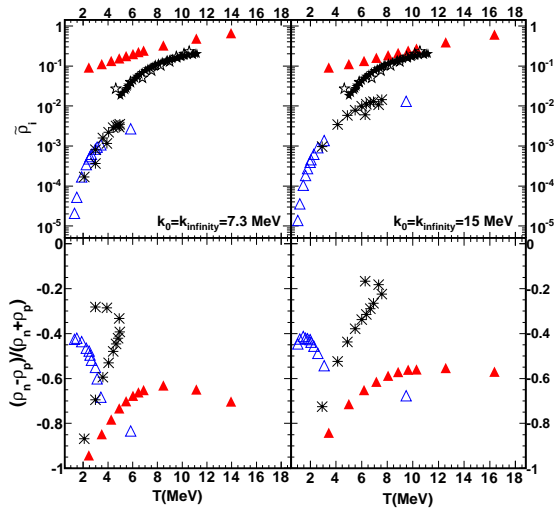}
\caption[]{(Color online) The total densities and density differences between neutrons and protons as function of temperature. (Top panels) The full stars are experimental results from \cite{qinprl12, hagelprl12, wadaprc12},  other symbols as in figs. \ref{Fig1} and \ref{Fig2}. (Bottom panels) The full triangles, open triangles and asterisks refer to QF, dth$\alpha$ and pnth DR respectively. }\label{Fig3}
\end{figure}
We also point out that the model assumes a barrier penetration of particles at various excitation energies \cite{geminicode, gemini1, gemini2, gemini3, gemini4, gemini5, gemini6, gemini7}. 
An effective radius is assumed for the system
$R=R_0+2.6fm=1.16A^{1/3}+2.6 fm$.
For a nucleus of mass $A=80$ this gives an effective radius $R=7.6 fm$ which is equivalent to a system having a density less than one third of the normal nuclear density. Of course this assumption might be in contrast with the density value obtained from the Fermi gas relation. However our goal is not to modify the model but to use it as a test bench, keeping in mind that these assumptions might be justified at low excitation energies (for which the model was proposed) and not at higher ones where fragmentation will dominate.
\begin{figure}
\centering
\includegraphics[width=1.0\columnwidth]{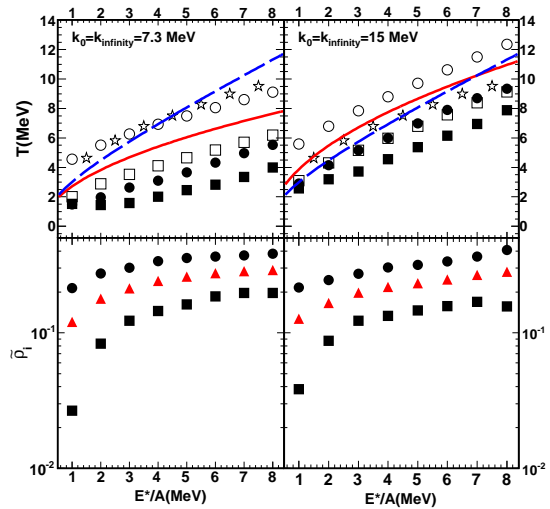}
\caption[]{(Color online) Same as fig.\ref{Fig1} for all steps model simulations.}\label{Fig4}
\end{figure}

Using the GEMINI code available in the literature \cite{geminicode, gemini1, gemini2, gemini3, gemini4, gemini5, gemini6, gemini7}, we have generated one million events for each excitation energy (or initial $T$).  
First we discuss the results for the simulations stopped at the first decay step where the relation of the excitation energy and temperature is given by eq. (\ref{energy}). In figure \ref{Fig1} we 
plot the temperature $T$ (top panels) and the density $\tilde \rho_i$ (bottom panels) as function of the excitation energy per particle of the intial hot system.
  Two different values of the $k_0$ Fermi gas parameter are used in the left and right hand side panels. The exact value of $T$
obtained from the Fermi gas relation, eq. (\ref{energy}), is given by the full (red) lines. Available experimental data from the current literature are given by the dashed (blue) lines \cite{elliott1, elliott2, elliott3} and
open stars \cite{justinT}, which we have reported for reference purposes only. 
The quantum fluctuation (QF) method, both for neutron and proton, agrees rather well with the exact result as expected since the basic assumption
in the method and in the GEMINI model is the same, i.e. a nucleus made of Fermions. 
The classical fluctuation method (CF) agrees with the exact method especially for the neutron case 
and at low excitation energies for both protons and neutrons \cite{magemini1, magemini2}. 
The reason for this behavior could be explained from the bottom part of figure \ref{Fig1}. The densities estimated only from the QF method  (the CF does not determine a density since the multiplicity fluctuation are equal to one classically)
are very low especially for neutrons. We expect that at low densities and relatively high $T$, classical and quantum methods should give similar results. As we will show the density obtained using the double ratio method is even smaller than
the one obtained from QF. In  figure \ref{Fig1}, the proton density is given by  full circles and the neutron density by  full squares, while the total density is given by  full triangle symbols. All densities have been
normalized to their respective ground state values, i.e. $\rho_{p0} = \rho_{n0}= 0.08 fm^{-3}$, $\rho_0 = 0.16 fm^{-3}$ and $\tilde\rho_i = \frac{\rho_i}{\rho_{i0}}$. The reason why we have extended the model to such high excitation energies where it is not necessarily justified, is because we wanted to show
that the estimated total density tends asymptotically to the value estimated from the Fermi gas relation, eq. (\ref{leveldens}), using the respective $k_0$ values which are
 given by the dotted horizontal lines in  figure \ref{Fig1}. At low excitation energies, different Q-values for particle emissions and barrier penetrations modify 
the fluctuations given by the Fermi gas entropy, eq. (\ref{entropy}), which results in lower densities as displayed. If these effects would be turned off, then fluctuations would 
arise from the Fermi gas entropy as for the quantum fluctuation method which is based on the same Fermi gas assumption \cite{landau}.

The discussion above can be extended to the double ratio (DR) method \cite{aldonc00, joe3, joeepja, albergot, msuld, msu1, msu2, qinprl12, hagelprl12, wadaprc12} and reported in figure \ref{Fig2}. The asterisks and crosses refer to the $T$ obtained from DR using deuteron, triton,$^3He$, $\alpha$ (dth$\alpha$)
and p, n, t, $^3He$ (pnth) combinations, respectively. The celebrated plateau of the caloric curve is observed in  figure \ref{Fig2} (top panels) 
especially for the dth$\alpha$ combination mostly used in the literature
\cite{joe3, pochodzalla1}. Notice the peculiar and maybe surprising result obtained using the pnth DR 
for which $T$ goes down for increasing excitation energy (top panels in figure \ref{Fig2}). A similar opposite behavior for the two cases is obtained
for the densities (bottom panels in figure \ref{Fig2}). In particular the pnth densities decrease for increasing excitation energy. All densities estimated from DR 
are much smaller than those  from QF and do not asymptotically tend to
the value obtained from the Fermi gas relation, eq. (\ref{leveldens}). Of course this is not surprising since we are using classical physics to estimate quantities obtained from a model based on a quantum (Fermi-gas)
system.
\begin{figure}
\centering
\includegraphics[width=1.0\columnwidth]{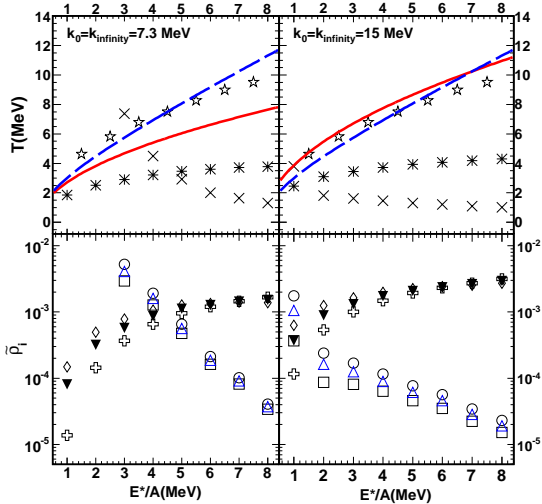}
\caption[]{(Color online) Same as fig. \ref{Fig2} for all steps model simulations.}\label{Fig5}
\end{figure}
\begin{figure}
\centering
\includegraphics[width=0.95\columnwidth]{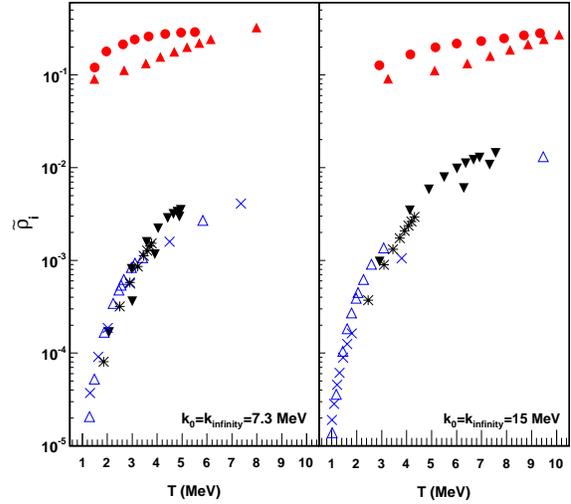}
\caption[]{(Color online) Total density vs temperature. Full circles (all steps) and full upward triangles (first step) from QF. Crosses (all steps) and open upward triangles (first step) from DR using pnth. Asterisks (all steps) and full downward triangles (first step) from DR using dht$\alpha$. }\label{Fig6}
\end{figure}

Another way to visualize the  results is by plotting density (top panels) and the difference of neutron and
 proton density (bottom panels) as function of $T$ as reported in figure \ref{Fig3}.  Now the surprising differences in the densities  obtained from 
the pnth and dth$\alpha$ cases
in figure \ref{Fig2} are not observed: the two DR methods agree which simply tells us that the control parameter is $T$ and not the excitation energy as it should be in a statistical model. The densities differ greatly in the QF and DR 
methods as observed before.  Equation (\ref{leveldens}) and 
the available data support higher densities \cite{qinprl12, hagelprl12, wadaprc12, justinT, ropke}. Furthermore the QF results tend asymptotically to the value expected from the Fermi gas and the used $k_0$ parameters.
Notice the large difference between  $n$ and $p$ densities as obtained in 
 different approaches (bottom panels in fig. \ref{Fig3}). In particular all different methods fail to reproduce the initial source value of zero $(N=Z)$ which should be recovered 
at high $T$. This is however a failure of the statistical model which we have pushed at high excitation energies where it is not justified. We notice that in a two-component system
 phase transition, the quantity plotted in the bottom part of figure \ref{Fig3},
could be considered as an order parameter \cite{aldoft, justinM}.

For completeness in figures \ref{Fig4} and \ref{Fig5} we display the results obtained when all steps in the statistical decay model are taken into account. 
These figures should be compared to figures \ref{Fig1} and \ref{Fig2} respectively.
We observe generally a decrease of $T$ and an increase of $\rho$ 
compared to the first step results.  This result implies that, if the general assumption of a sequential decay is correct,
then the derived $T$ and $\rho$ estimated in the different methods are effective values influenced by the secondary decay.  Within the fluctuation method, it seems that the initial $T$ is somehow between
the classical and the quantum cases, while the DR method fail in all cases. However, as we have seen in figure \ref{Fig3}, plots of $\rho$ and $T$ as function of excitation energy might be misleading as
in the pnth and dth$\alpha$ cases. In figure \ref{Fig6} we plot $\rho$ as function of $T$ obtained from  different assumptions both at the first decay and all decay steps.  As we see in the figure all results roughly 
collapse  in single curves, especially results from the DR method,
 which suggests that indeed the values of $T$ might shift down due to the secondary decay, 
however the corresponding density is also modified in such a way to collapse in a single curve.
This result should be compared to similar calculations using CoMD, see fig. (26) in ref. \cite{huappnp}.

In conclusion, in this paper we have compared different proposed methods to extract density and temperature using a statistical sequential model. We have shown that
the model observables are better reproduced by the quantum fluctuations method since the same physical ingredient, the Fermi gas, is used.  Double ratios fail because of the classical
assumptions as it should be. However, the feature that different ratios give different $T$ and $\rho$ as function of excitation energy is misleading. 
An agreement of the different particle ratios
is observed when the temperature is used as a control parameter as it should be in a statistical environment. Secondary decays support again the QF method as compared to the DR and differences might be highlighted  by
plotting densities as function of the control parameter $T$.
 
 \section*{Acknowledgement}
We thank prof. J. Natowitz for stimulating discussions.


\begin{thebibliography}{99}
\bibitem{baoanrep08} B.A. Li, L.W. Chen and C.M. Ko, \Journal{\PREP}{464}{113}{2008}
\bibitem{huappnp}G. Giuliani, H. Zheng and A. Bonasera, \Journal{\PROG}{76}{116}{2014}
\bibitem{baranrep05}V. Baran, M. Colonna, V. Greco and M. Di Toro, \Journal{\PREP}{410}{335}{2005}
\bibitem{skyrme3} A.W. Steiner, M. Prakash, J.M. Lattimer and P.J. Ellis, \Journal{\PREP}{411}{325}{2005}
\bibitem{nsob12} J.M. Lattimer and M. Prakash, \Journal{\PREP}{442}{109}{2007}
\bibitem{aldonc00} A. Bonasera, M. Bruno, C.O. Dorso and P.F. Mastinu, \Journal{\NC}{23}{2}{2000}
\bibitem{joe3} J.B. Natowitz {\it et al.}, \Journal{\PRC}{65}{034618}{2002}
\bibitem{joeepja} A. Keli\'c, J.B. Natowitz and K.H. Schmidt, \Journal{\EPJA}{30}{203}{2006}
\bibitem{msuld}M.B. Tsang, F. Zhu, W.G. Lynch, A. Aranda, D.R. Bowman, R.T. de Souza, C.K. Gelbke, Y.D. Kim, L. Phair, S. Pratt, C. Williams, H.M. Xu and W.A. Friedman, \Journal{\PRC}{53}{R1057}{1996}
\bibitem {albergot}  S. Albergo, S. Costa, E. Costanzo and A. Rubbino, \Journal{\NCA}{89}{1}{1985}
\bibitem {msu1} S. Das Gupta, J. Pan and M.B. Tsang, \Journal{\PRC}{54}{R2820}{1996}
\bibitem {msu2} H. Xi, W.G. Lynch, M.B. Tsang, W.A. Friedman and D. Durand, \Journal{\PRC}{59}{1567}{1999}
\bibitem{qinprl12}L. Qin, K. Hagel, R. Wada, J.B. Natowitz, S. Shlomo, A. Bonasera, G. R\"opke, S. Typel, Z. Chen, M. Huang, J. Wang, H. Zheng, S. Kowalski, M. Barbui, M.R.D. Rodrigues, K. Schmidt, D. Fabris, M. Lunardon, S. Moretto, G. Nebbia, S. Pesente, V. Rizzi, G. Viesti, M. Cinausero, G. Prete, T. Keutgen, Y. EI Masri, Z. Majka and Y.G. Ma, \Journal{\PRL}{108}{172701}{2012}
\bibitem{hagelprl12}K. Hagel, R. Wada, L. Qin, J.B. Natowitz, S. Shlomo, A. Bonasera, G. R\"opke, S. Typel, Z. Chen, M. Huang, J. Wang, H. Zheng, S. Kowalski, C. Bottosso, M. Barbui, M.R.D. Rodrigues, K. Schmidt, D. Fabris, M. Lunardon, S. Moretto, G. Nebbia, S. Pesente, V. Rizzi, G. Viesti, M. Cinausero, G. Prete, T. Keutgen, Y. EI Masri and Z. Majka, \Journal{\PRL}{108}{062702}{2012}
\bibitem{wadaprc12} R. Wada, K. Hagel, L. Qin, J.B. Natowitz, Y.G. Ma, G. R\"opke, S. Shlomo, A. Bonasera, S. Typel, Z. Chen, M. Huang, J. Wang, H. Zheng, S. Kowalski, C. Bottosso, M. Barbui, M.R.D. Rodrigues, K. Schmidt, D. Fabris, M. Lunardon, S. Moretto, G. Nebbia, S. Pesente, V. Rizzi, G. Viesti, M. Cinausero, G. Prete, T. Keutgen, Y. EI Masri and Z. Majka, \Journal{\PRC}{85}{064618}{2012}
\bibitem{mekjian1}A. Mekjian, \Journal{\PRL}{38}{640}{1977}
\bibitem{mekjian2}A.Z. Mekjian, \Journal{\PRC}{17}{1051}{1978}
\bibitem{awes}T.C. Awes, G. Poggi, C.K. Gelbke, B.B. Back, B.G. Glagola, H. Breuer and V.E. Viola, Jr, \Journal{\PRC}{24}{89}{1981}
\bibitem{csernai}L.P. Csernai and J.I. Kapusta, \Journal{\PREP}{131}{223}{1986}
\bibitem{ropke}G. R\"opke, S. Shlomo, A. Bonasera, J.B. Natowitz, S.J. Yennello, A.B. McIntosh, J. Mabiala, L. Qin, S. Kowalski, K. Hagel, M. Barbui, K. Schmidt, G. Giulani, H. Zheng and S. Wuenschel, \Journal{\PRC}{88}{024609}{2013}
\bibitem{wuenschel1} S. Wuenschel {\it et al.}, \Journal{\NPA}{843}{1}{2010}
\bibitem{hua3}H. Zheng and A. Bonasera, \Journal{\PLB}{696}{178}{2011}
\bibitem{hua4}H. Zheng and A. Bonasera, \Journal{\PRC}{86}{027602}{2012}
\bibitem{hua5}H. Zheng, G. Giuliani and A. Bonasera, \Journal{\NPA}{892}{43}{2012}
\bibitem{hua6}H. Zheng, G. Giuliani and A. Bonasera, \Journal{\PRC}{88}{024607}{2013}
\bibitem{hua7}H. Zheng, G. Giuliani and A. Bonasera, \Journal{\JPG}{41}{055109}{2014}
\bibitem{comd1}M. Papa, T. Maruyama and A. Bonasera, \Journal{\PRC}{64}{024612}{2001}
\bibitem{comd2}M. Papa, G. Giuliani and A. Bonasera, \Journal{\JCP}{208}{403}{2005}
\bibitem{justinT} J. Mabiala, A. Bonasera, H. Zheng, A.B. McIntosh, Z. Kohley, P. Cammarata, K. Hagel, L. Heilborn, L.W. May, A. Raphelt, G.A. Souliotis, A. Zarrella and S.J. Yennello, \Journal{\IJMPE}{Vol. 22, No. 12}{1350090}{2013}
\bibitem{brianT} B.C. Stein, A. Bonasera, G.A. Souliotis, H. Zheng, P.J. Cammarata, A.J. Echeverria, L. Heilborn, A.L. Keksis, Z. Kohley, J. Mabiala, P. Marini, L.W. May, A.B. McIntosh, C. Richers, D.V. shetty, S.N. Soisson, R. Tripathi, S. Wuenschel and S.J. Yennello, \Journal{\JPG}{41}{025108}{2014}
\bibitem{esteve1} J. Esteve {\it et al.}, \Journal{\PRL}{96}{130403}{2006}
\bibitem{muller1} T. M\"uller {\it et al.}, \Journal{\PRL}{105}{040401}{2010}
\bibitem{sanner1} C. Sanner {\it et al.}, \Journal{\PRL}{105}{040402}{2010}
\bibitem{westbrook1} C.I. Westbrook, \Journal{\PHY}{3}{59}{2010}
\bibitem{palao}P. Marini {\it et al.} in preparation. 
\bibitem{geminicode}R. J. Charity, computer code GEMINI, see http://www.chemistry.wustl.edu/$\sim$rc
\bibitem{gemini1} J.O. Newton, D.J. Hinde, R.J. Charity, J.R. Leigh, J.J.M. Bokhorst, A. Chatterjee, G.S. Foote and S. Ogaza, \Journal{\NPA}{483}{126}{1988}
\bibitem{gemini2} R.J. Charity, M.A. McMahan, G.J. Wozniak, R.J. McDonald, L.G. Moretto, D.G. Sarantites, L.G. Sobotka, G. Guarino, A. Panteleo, L. Fiore, A. Gobbi and K. Hildenbrand, \Journal{\NPA}{483}{371}{1988}
\bibitem{gemini3} D.R. Bowman, G.F. Peaslee, N. Colonna, R.J. Charity, M.A. McMahan, D. Delis, H. Han, K. Jing, G.J. Wozniak, L.G. Moretto, W.L. Kehoe, B. Libby, A.C. Mignerey, A. Moroni, S. Angius, I. Iori, A. Pantaleo and G. Guarino, \Journal{\NPA}{523}{386}{1991}
\bibitem{gemini4} R.J. Charity, \Journal{\PRC}{53}{512}{1996}
\bibitem{gemini5} R.J. Charity, \Journal{\PRC}{61}{054614}{2000}
\bibitem{gemini6} L.G. Sobotka, R.J. Charity, J.T\~oke and W.U. Schr\"oder, \Journal{\PRL}{93}{132702}{2004}
\bibitem{gemini7} R.J. Charity, \Journal{\PRC}{82}{014610}{2010}
\bibitem{magemini1} W.D. Tian, Y.G. Ma, X.Z. Cai, D.Q. Fang, W. Guo, W.Q. Shen, K. Wang, H.W. Wang and M. Veselsky, \Journal{\PRC}{76}{024607}{2007}
\bibitem{magemini2} P. Zhou, W.D. Tian, Y.G. Ma, X.Z. Cai, D.Q. Fang, and H.W. Wang, \Journal{\PRC}{84}{037605}{2011}
\bibitem{landau} L. Landau and F. Lifshits, Statistical Physics (Pergamon, New York) 1980. 
\bibitem{huang} K. Huang, Statistical Mechanics (J. Wiley and Sons, New York) 1987,  second ed..
\bibitem{elliott1} L.G. Moretto, J.B. Elliott, L. Phair and P.T. Lake, \Journal{\JPG}{38}{113101}{2011}
\bibitem{elliott2} J.B. Elliott, P.T. Lake, L.G. Moretto and L. Phair, \Journal{\PRC}{87}{054622}{2013}
\bibitem{elliott3} J.B. Elliott {\it et al.}, \Journal{\PRC}{67}{024609}{2003}
\bibitem{pochodzalla1} J. Pochodzalla {\it et al.}, \Journal{\PRL}{75}{1040}{1995}
\bibitem{aldoft}A. Bonasera, Z. Chen, R. Wada, K. Hagel, J. Natowitz, P. Sahu, L. Qin, S. Kowalski, T. Keutgen, T. Materna and T. Nakagawa, \Journal{\PRL}{101}{122702}{2008}
\bibitem{justinM} J. Mabiala {\it et al.} in preparation.



%\bibitem{pathria} R. K. Pathria, Statistical Mechanics (Elsevier Pte Ltd, Singapore) 2003, second ed..
%\bibitem{sachie} S. Kimura and A. Bonasera, Phys. Rev. A \textbf{72}, 014703 (2005).
%\bibitem{hua} H. Zheng and A. Bonasera, Phys. Rev. C \textbf{83}, 057602 (2011).
%\bibitem{he41}H. A. Kierstead, Phys. Rev \textbf{Vol 162}, 153 (1967).
%\bibitem{he42} S. M. Aperko, Phys. Rev. B \textbf{Vol 60}, 3052 (1999).
%\bibitem{he43} G. H. Bauer, D. M. Ceperley and N. Goldenfeld, Phys. Rev. B \textbf{Vol 61}, 9055 (2000).
%\bibitem{he44} S. Balibar, H. Alles and A. Y. Parshin, Rev of Mod Phys \textbf{Vol 77}, 317 (2005).
\end{thebibliography}
\end{document}